\documentclass{article}
\usepackage{spconfa4,amsmath,amssymb,graphicx}
\usepackage{tikz}
\usetikzlibrary{positioning}
\usepackage{booktabs}  
\usepackage{float}     
\usepackage{array}   
\usepackage{placeins}
\usepackage{subcaption}
\usepackage{multirow} %
\usepackage{cite}
\usepackage{hyperref}
\usepackage{balance}
\usepackage{etoolbox}
\apptocmd{\thebibliography}{\interlinepenalty=10000\relax}{}{}

\title{MASK-BASED SPEECH ENHANCEMENT FOR SPATIAL AUDIO: EFFECTS OF SIGNAL REPRESENTATION AND CHANNEL-SPECIFIC MASKING}

\twoauthors
  {Sheli Hendel and Boaz Rafaely}
	{Ben-Gurion University of the Negev\\
	School of Electrical and Computer Engineering\\
	Beer-Sheva 84105, Israel}
  {Dorothea Kolossa}
	{Technische Universität Berlin\\
    Human-centered Information Technologies\\
	Berlin 10623, Germany}

\begin{document}
\maketitle
\thispagestyle{empty}
\begingroup\renewcommand{\thefootnote}{}
\footnotetext{Funded by Deutsche Forschungsgemeinschaft (DFG, German Research Foundation) – Project number KO3434/9-1.}

\begin{abstract}
Speech enhancement for spatial audio needs to improve speech quality and intelligibility while preserving the spatial characteristics of the acoustic scene. Time-frequency masking provides an effective approach for speech enhancement, but its performance and spatial fidelity may depend strongly on the signal representation in which the mask is applied. This work systematically compares time-frequency masking in the microphone, beamformer, and first-order Ambisonics domains in simulated reverberant acoustic scenes. Both an ideal ratio mask (IRM) and a lightweight deep-learning-based mask estimator are considered, with performance evaluated in terms of speech reconstruction, perceptual quality, intelligibility, reverberation preservation, and binaural spatial-cue preservation. The results show that channel-specific masking in the beamformer domain provides the highest speech quality and intelligibility, while channel-specific masking in the Ambisonics domain better preserves the spatial characteristics of the residual interference. Reducing the spatial representation from third- to first-order Ambisonics substantially reduces the performance gap between the beamformer and Ambisonics domains due to the lower spatial resolution of the resulting beamformers. Furthermore, a shared mask across first-order Ambisonics channels achieves speech enhancement performance close to that of channel-specific masking while requiring only a single mask-estimation network. These results demonstrate that the choice of signal representation and masking strategy should be considered jointly, and indicate that first-order Ambisonics with shared masking provides a practical compromise between speech enhancement performance, spatial fidelity, and computational complexity.

\end{abstract}
\section{Introduction}
\label{sec:intro}

Speech enhancement aims to improve speech quality and intelligibility in noisy environments by suppressing interfering sounds. Time-frequency masking has become one of the most widely used approaches, particularly when the masks are estimated using deep neural networks trained to separate speech from background noise~\cite{narayanan2013ideal,wang2018supervised,zheng2023sixty,wang2014training,huang2014deep,huang2015joint}. These methods have demonstrated substantial improvements over classical signal processing techniques and are typically optimized and evaluated using objectives related to speech quality or intelligibility, such as scale-invariant signal-to-distortion ratio (SI-SDR), Perceptual Evaluation of Speech Quality (PESQ), and Short-Time Objective Intelligibility (STOI)~\cite{5713237,le2019sdr,rix2001perceptual,luo2019conv,huang2015joint,kolbaek2020loss,fu2018end}. Although multichannel recordings are often used as input, the enhanced output is typically a single-channel speech signal, and the learning objective focuses primarily on suppressing interference while maximizing speech quality and intelligibility.

While speech quality and intelligibility is important for speech communication, the increasing use of immersive audio technologies, including augmented and virtual reality (AR and VR), hearing assistance and three-dimensional audio reproduction, has motivated the extension of speech enhancement methods to multichannel spatial audio~\cite{herzog2023ambisep,wang2025lightweight,leng2021compromise,sun2019deep,kumar2025relative}. In these applications, enhancement algorithms aim to preserve not only speech quality and intelligibility but also the spatial information required for accurate localization, scene awareness, listener comfort, and spatial release from masking (SRM)~\cite{blauert2013technology,hauth2020modeling,lugasi2020speech,rafaely2022spatial}.  However, masking may distort inter-channel spatial information, thereby altering spatial perception. While previous studies have demonstrated the feasibility of applying speech enhancement within the context of spatial audio systems, a gap remains in the understanding of the effective approach for achieving both a high level of speech enhancement and high-quality spatial reproduction, in particular for masking-based methods.

Recent work has begun to address this gap. In~\cite{lugasi2020speech}, masking in the beamformer and Ambisonics domains was compared, showing that the latter better preserves the spatial information of both the target speech and the residual disturbance. More recently,~\cite{hendel2026maskbasedspeechenhancementspatial} systematically compared the time-frequency ideal ratio mask (IRM) applied to microphone, beamformer, and Ambisonics representations, demonstrating distinct trade-offs between speech enhancement performance and spatial fidelity. AmbiSep~\cite{herzog2023ambisep,Herzog2022AmbiSep} compared the performance of Ambisonics and beamformers channels with transformer-based neural-network architecture to jointly perform speech separation and noise reduction, while preserving spatial information. However, the analysis did not include microphone channels nor spatial measures such as preservation of reverberation, and interaural time difference (ITD) and interaural level difference (ILD) computed using reproduced binaural signals. 
Despite these advances, a gap remains: no comprehensive analysis yet provides both insight and practical design guidance for achieving high speech enhancement alongside spatial fidelity, including preservation of perceived reverberation and localization of both target and interfering sources, across oracle and practical neural-network-based masking systems.

This work addresses this gap by providing a systematic comparison of oracle and deep learning-based time–frequency masking across signal representations, with the aim of deriving both insight and practical design guidance for spatial-audio based speech enhancement systems.
Specifically, we compare time-frequency masks estimated by a lightweight network and oracle IRM applied in three signal representations: microphone signals, beamformer outputs, and first-order Ambisonics. Focusing on first-order Ambisonics enables evaluation under realistic microphone-array constraints encountered in practical immersive audio systems. The methods are evaluated using simulated reverberant acoustic scenes with respect to speech enhancement performance, perceptual quality, intelligibility, reverberation preservation, and binaural cue preservation.

The main contributions of this work are summarized as follows: (i) showing that channel-specific masking applied to beamformer channels best preserves speech quality and intelligibility compared to Ambisonics and microphone channels; (ii) showing that channel-specific masking applied to Ambisonics channels best preserves the spatial attributes of the residual disturbance compared to beamformer and microphone channels; and (iii) showing that a shared mask applied across Ambisonics channels provides speech quality and intelligibility close to that of channel-specific masking, with only a slight performance reduction, therefore offering a more efficient implementation.

\section{BACKGROUND}
\label{sec:background}
\subsection{Signal Model}
Consider a reverberant acoustic environment comprising a target speech source, interfering sources, and stationary background noise. The acoustic scene is observed using a rigid spherical microphone array with $Q$ microphones. In the frequency domain, the microphone signals are modeled as

\begin{equation}
\mathbf{x}(f)=\mathbf{x}^{\mathrm{dir}}(f)+\mathbf{x}^{\mathrm{rev}}(f)+\mathbf{n}(f),
\label{eq:mic}
\end{equation}
where $\mathbf{x}(f)\in\mathbb{C}^{Q\times1}$ denotes the microphone signal vector at frequency $f$. The microphones are located on a sphere of radius $r$ at positions $\Phi_q=(\theta_q,\phi_q)$, $q=1,\ldots,Q$. The terms $\mathbf{x}^{\mathrm{dir}}(f)$ and $\mathbf{x}^{\mathrm{rev}}(f)$ correspond to the direct-path and reverberant components of the target speech, respectively, while $\mathbf{n}(f)$ represents the combined contribution of background noise and interfering sources.

The same sound field can be represented in the spherical harmonics (SH), or Ambisonics, domain~\cite{lugasi2020speech} as

\begin{equation}
\mathbf{a}_{nm}(f)
=
\mathbf{a}_{nm}^{\mathrm{dir}}(f)
+
\mathbf{a}_{nm}^{\mathrm{rev}}(f)
+
\mathbf{a}_{nm}^{\mathrm{noise}}(f),
\label{eq:Ambisonics}
\end{equation}
where $\mathbf{a}_{nm}(f)=[a_{00}(f),a_{1(-1)}(f),a_{10}(f),\ldots,a_{NN}(f)]^{T}$ $\in\mathbb{C}^{(N+1)^2\times1}$ is the vector of the Ambisonics coefficients of order $N$~\cite{rafaely2015fundamentals}. The vectors $\mathbf{a}_{nm}^{\mathrm{dir}}(f)$, $\mathbf{a}_{nm}^{\mathrm{rev}}(f)$, and $\mathbf{a}_{nm}^{\mathrm{noise}}(f)$ denote the direct, reverberant, and noise components, respectively.

Finally, the acoustic scene can be represented in the beamformer domain by applying a set of spatial beamformers:
\begin{equation}
\mathbf{b}(f)
=
\mathbf{b}^{\mathrm{dir}}(f)
+
\mathbf{b}^{\mathrm{rev}}(f)
+
\mathbf{b}^{\mathrm{noise}}(f),
\label{eq:pwd_stft}
\end{equation}
where $\mathbf{b}(f)\in\mathbb{C}^{B\times1}$ contains the outputs of $B$ beamformers steered toward directions represented in spherical coordinates $\Psi_b=(\theta_b,\phi_b)$, $b=1,\ldots,B$. The vectors $\mathbf{b}^{\mathrm{dir}}(f)$, $\mathbf{b}^{\mathrm{rev}}(f)$, and $\mathbf{b}^{\mathrm{noise}}(f)$ represent the corresponding direct-path, reverberant, and noise contributions.

\subsection{Transformation Between Signal Representations}
Throughout this work, the Ambisonics representation in Eq.~\eqref{eq:Ambisonics} serves as the reference signal representation. The signals in the microphone-domain and the beamformer-domain are subsequently derived from the Ambisonics coefficients following the formulation in~\cite{rafaely2015fundamentals}. In order to transform from the SH domain to the beamformer or microphone domain, we use:
\begin{align}
\mathbf{b}(f) &= \mathbf{Y}_{\text{Beam}}(\mathbf{\Psi})\mathbf{a}_{nm}(f) \label{eq:decoder_Beam} \\
\mathbf{x}(f) &= \mathbf{Y}_{\text{Mic}}(\mathbf{\Phi})\mathbf{B}(fr)\mathbf{a}_{nm}(f)  \label{eq:decoder_mic}
\end{align}
where $\mathbf{Y}_{\text{Beam}}(\Psi)$ and $\mathbf{Y}_{\text{Mic}}(\Phi)$ are $B\times(N+1)^2$ and $Q\times(N+1)^2$ matrices of SH as in~\cite{lugasi2020speech,rafaely2015fundamentals}. $\mathbf{\Psi}=[\Psi_1,\Psi_2,...,\Psi_B]$ is the vector of the $B$ look directions and $\mathbf{\Phi}=[\Phi_1,\Phi_2,...,\Phi_Q]$ is the vector of the $Q$ microphones locations on a sphere with radius $r$. $\mathbf{B}(fr)=\text{diag}(b_0(fr),b_1(fr),b_1(fr),b_1(fr),...,b_N(fr))$ is a $(N+1)^2\times(N+1)^2$ diagonal matrix where $b_n(fr)$ represents the radial functions that depend on the frequency and distance from the origin. To transform back to the SH domain we will use:
\begin{align}
\mathbf{a}_{nm}(f) &= \mathbf{Y}_{\text{Beam}}^{\dagger}(\mathbf{\Psi})\mathbf{b}(f) \label{eq:encoder_Beam} \\
\mathbf{a}_{nm}(f) &= (\mathbf{Y}_{\text{Mic}}(\mathbf{\Phi})\mathbf{B}(fr))^\dagger \mathbf{x}(f)
\label{eq:encoder_mic}
\end{align}
where  $(.)^\dagger$ is pseudo-inverse, assuming that $Q, B\geq (N+1)^2$.
Because we use pseudo-inverse, Eq.~\eqref{eq:encoder_mic} may introduce an error which may be small if the $B$ and $Q$ directions sample the sphere nearly uniformly \cite{rafaely2015fundamentals}.

\subsection{Time-Frequency Masks}
The frequency-domain signals are transformed to the time domain and then to the time-frequency domain using the short-time Fourier transform (STFT), yielding the representations $\mathbf{x}(\tau,\nu)$, $\mathbf{a}_{nm}(\tau,\nu)$, and $\mathbf{b}(\tau,\nu)$. The desired signal component is then estimated by applying a time-frequency mask to each representation:
\begin{align}
\hat{\mathbf{x}}(\tau,\nu) &= \mathbf{M}_{\text{Mic}}(\tau,\nu) \mathbf{x}(\tau,\nu) \label{eq:mask_mic} \\
\hat{\mathbf{a}}_{nm}(\tau,\nu) &= \mathbf{M}_{\text{Amb}}(\tau,\nu) \mathbf{a}_{nm}(\tau,\nu) \label{eq:mask_amb} \\
\hat{\mathbf{b}}(\tau,\nu) &= \mathbf{M}_{\text{Beam}}(\tau,\nu) \mathbf{b}(\tau,\nu) \label{eq:mask_Beam}
\end{align}
where $\mathbf{M}_{\text{Mic}}(\tau,\nu)$, $\mathbf{M}_{\text{Amb}}(\tau,\nu)$ and $\mathbf{M}_{\text{Beam}}(\tau,\nu)$ are diagonal matrices. Here the desired signal is defined as the reverberant target. At the end of this process,
the signals are transformed back to the time domain. In this work, we will focus on two types of masks:

\emph{1) Ideal ratio mask (IRM):}
We will use the IRM as in~\cite{hendel2026maskbasedspeechenhancementspatial,lugasi2020speech}:
\begin{equation}
M(\text{ch},\tau,\nu) =  \frac{SNR(\text{ch},\tau,\nu)}{SNR(\text{ch},\tau,\nu)+1}     
\label{eq:IRM}
\end{equation}
\begin{equation}
\mathbf{M}(\tau,\nu) =   \operatorname{diag}(M(0,\tau,\nu), M(1,\tau,\nu),...)
\end{equation}
where $SNR(\text{ch},\tau,\nu)$ is the ratio of the power of the desired signal to that of the undesired signal for each channel. 

\emph{2) FT-JNF estimated mask:} The FT-JNF (frequency--temporal joint non-linear filter) as in~\cite{tesch2022insights} is a data-driven model that estimates the mask directly from the noisy multichannel input.

\subsection{Binaural Signal Representation}
The enhanced signals are rendered as binaural signals by applying an HRTF-based Ambisonics decoder in the frequency domain. The binaural left/right ear signals are obtained as
\begin{equation}
p^{l/r}(f) = \tilde{\mathbf{h}}^{l/r}_{nm}(f)^T \mathbf{a}_{nm}(f),
\label{eq:Binaural}
\end{equation}
where $\tilde{\mathbf{h}}^{l/r}_{nm}(f)$ denotes the modified SH-domain representation of the HRTF. It is derived from the original SH-domain HRTF coefficients $\mathbf{h}^{l/r}_{nm}(f)$ according to $\tilde{h}^{l/r}_{nm}(f)=(-1)^m h^{l/r}_{n(-m)}(f)$~\cite{gayer2025ambisonics}. When the enhanced signal is obtained in the microphone or beamformer domains, it is first converted back to the SH domain using Eqs.~\eqref{eq:encoder_Beam} and~\eqref{eq:encoder_mic}, after which the binaural decoding is performed.

\section{Proposed Analysis}
\label{sec:proposed_analysis}
The central objective of this work is to systematically analyze the
interaction between signal representation and channel-specific
masking application in spatial-audio speech enhancement. Rather
than evaluating speech enhancement independently of spatial
reproduction, the proposed analysis examines the trade-off between
suppressing undesired signals and preserving the spatial
characteristics of the acoustic scene.

\subsection{Research Questions}
This objective is addressed through these research questions:
\begin{enumerate}
    \item How does the signal representation affect the trade-off between
    speech enhancement and spatial fidelity?
    
    \item To what extent does channel-specific masking improve
    enhancement compared with a computationally-efficient shared mask, and does this effect
    depend on the signal representation?
\end{enumerate}

\subsection{Analysis Dimensions: Signal Representation and Masking Strategy}
To answer these questions, the analysis is built around two
dimensions.
The first dimension is signal representation. Three representations
are considered: microphone channels, first-order Ambisonics
channels, and beamformer outputs. The key premise is that masking
may affect these representations differently, because their channels
encode spatial information in different ways. 
The analysis additionally compares first- and third-order Ambisonics representations, to investigate the effect of Ambisonics order on all three representations. 
The second dimension is masking strategy. A shared mask uses a
single mask applied to all channels. Channel-specific masking instead
estimates an independent mask for each channel. Channel-specific
masks may improve speech enhancement over a shared mask by adapting to the characteristics of each channel, but can also introduce channel-dependent modifications that distort spatial cues. 

\subsection{Oracle vs.\ Estimated Masking}
To separate the effect of masking strategy from the effect of mask
estimation itself, both the ideal ratio mask (IRM) and the lightweight
FT-JNF neural estimator are considered. The IRM provides an oracle
reference for achievable performance, while FT-JNF serves as a
practical mask estimator, used to determine whether the relationships
observed with oracle masking remain when the mask must be
estimated from the noisy multichannel signal.

\subsection{Evaluation Framework}
Finally, the analysis jointly considers speech quality and intelligibility, reverberation preservation, and
binaural cues of both the target and residual interference. This
allows the trade-off between enhancement performance and spatial
fidelity to be evaluated across signal representations and masking
strategies. The resulting framework is evaluated under controlled
acoustic conditions in the simulation study that follows.

\section{OBJECTIVE MEASURES OF PERFORMANCE}
\label{sec:MEASURES}
Performance is assessed using objective metrics that, as mentioned above, evaluate speech intelligibility, speech quality, binaural cues and reverberation preservation. These are detailed below. 

\begin{itemize}

\item \textbf{SI-SDR}: Signal reconstruction accuracy is quantified using SI-SDR~\cite{le2019sdr}. The metric is calculated independently for the left and right binaural channels and averaged to obtain the overall performance measure. The reverberant target speech signal is used as the reference.

\item $\mathbf{\Delta}$\textbf{ITD},\, $\mathbf{\Delta}$\textbf{ILD}:
The preservation of binaural spatial information is evaluated through the ITD and ILD~\cite{blauert2013technology,hauth2020modeling}. ITD is the dominant localization cue below approximately 1.5~kHz, whereas ILD dominates at higher frequencies~\cite{blauert2013technology,hauth2020modeling}. These measures are computed using the direct-path component of the target signal. The spatial distortion introduced by the enhancement process is measured as the absolute difference between the estimated and reference ITD and ILD values, denoted as $\Delta\mathrm{ITD}$ and $\Delta\mathrm{ILD}$.

\item $\mathbf{\Delta}$\textbf{C}$_\mathbf{50}$:
The influence of the enhancement process on reverberation is evaluated using the clarity index $\mathrm{C}_{50}$~\cite{christensen2013iso}. This measure quantifies the ratio between the early arriving energy within the first 50~ms and the subsequent late reverberation energy. The deviation from the reference reverberant target signal is calculated as the absolute $\mathrm{C}_{50}$ error, averaged over the two ears, and denoted as $\Delta\mathrm{C}_{50}$.

\item \textbf{DBSTOI}:
Speech intelligibility is measured using the deterministic binaural STOI (DBSTOI)~\cite{andersen2016method}, which accounts for binaural effects and non-linear auditory processing. The direct-path target speech serves as the reference for this evaluation. 

\item \textbf{PESQ}: Speech quality is assessed using PESQ~\cite{rix2001perceptual}, where the reverberant target signal is used as the reference. PESQ scores are computed separately for each ear and averaged to obtain the final score.
\end{itemize}

\section{Simulation Study}
\label{sec:ANALYSIS}
This section describes the simulation framework used to evaluate the considered speech enhancement methods. The experiments are designed to provide a controlled comparison of microphone, Ambisonics, and beamformer representations under identical acoustic conditions. The simulated scenes include reverberation, directional interference, and background noise, allowing both speech enhancement performance and the preservation of spatial information to be evaluated. The simulation setup, signal representations, and training procedure are described first, followed by the experimental results.

\subsection{Setup}
We generate a simulated dataset using \texttt{shroom}~\cite{gayer2026shroom}, which provides an implementation of the source-image model~\cite{allen1979image}. Each scene contains a target speaker, an interfering speaker, and a stationary background-noise source, captured by a virtual receiver. The dataset consists of a total of 600 examples for training, 200 examples for validation, and 100 examples for testing, with each example of 6 seconds duration.

For every scene, a rectangular room is randomly generated with varying dimensions and wall absorption, while the receiver is placed at a random position at least $1$~m from the walls. The target speaker is positioned directly in front of the receiver at a random distance, and the interfering speaker is placed at the same distance but at a random direction. A stationary babble-noise source is placed in a randomly selected room corner to emulate a highly reverberant noise field. The interferer and noise source signal are scaled according to specified SNR conditions, and the total undesired signal is the sum of the interfering speaker and the noise. Table~\ref{tab:simulation_data} details how source positions and other scene parameters are selected.

Each scene is simulated at a high spherical-harmonics order with $N=10$, from which all signal representations are derived so that the three domains describe the same underlying sound field. For the main experiments, the Ambisonics signals are then truncated to first-order. The microphone signals are rendered from the high-order field using a rigid spherical array of $Q=4$ microphones with radius $r=0.042$~m, while the beamformer signals are obtained by a signal-independent plane-wave decomposition of the first-order field onto $B=4$ look directions \cite{rafaely2015fundamentals}. The microphone locations and beamformer look directions are distributed uniformly on the sphere, with the first direction aligned to the target. For the third-order Ambisonics channels comparison,
the corresponding representations are instead constructed at order $N=3$, using $Q=24$ microphones and $B=24$ beamformer look directions, with the latter distributed using a near-uniform Fibonacci grid. Channel types are denoted as follows:
\begin{itemize}
    \item \textbf{TFM}: Time-frequency microphones
    \item \textbf{TFA}: Time-frequency Ambisonics
    \item \textbf{TFB}: Time-frequency Beamforming
\end{itemize}

\emph{Training and validation data:} To expose the network to a wide variety of spatial configurations, the interferer azimuth and height are randomly sampled according to the distributions specified in Table~\ref{tab:simulation_data}.

\emph{Testing data:} For the evaluation set, the interferer is restricted to two fixed azimuth directions relative to the target, providing a large spatial separation between the target and interference. This makes the preservation of the interference spatial cues a perceptually meaningful task. For each test scene, additional reference signals are stored: the direct-sound from the target ($\mathbf{a}^{\text{dir}}_{nm}$), the direct-sound from the interference, and the target signal convolved with the room ARIR truncated to $50$~ms after the direct arrival, used to compute $\text{C}_{50}$.

\subsection{Methodology}
Two types of speech enhancement masks are computed as outlined in Sec. \ref{sec:proposed_analysis}. The IRM mask is computed directly from the known desired and undesired components, as in~\cite{hendel2026maskbasedspeechenhancementspatial,lugasi2020speech}. 
For the data driven method, we train a bidirectional LSTM (FT-JNF) as in~\cite{tesch2022insights} that estimates a complex time-frequency mask from the noisy signal. The network input is the one-sided STFT of the noisy multichannel mixture in a given domain, with the real and imaginary parts of all channels concatenated to form a $[\,T\times F\times 2C\,]$ tensor, where $C=4$ is the number of channels in that domain. The estimated mask is applied to the STFT of a chosen reference channel, and the masked STFT is transformed back to the time domain. Unlike~\cite{tesch2022insights}, here the network is trained to maximize the signal-to-distortion ratio (SDR) between the reconstructed and the desired reference signal, and not SI-SDR. We chose to use SDR to preserve the scale between the channels, which is important for the spatial processing that follow. The reverberant target is used as the desired reference so that the network preserves target signal reverberation \cite{hendel2026maskbasedspeechenhancementspatial}. The FT-JNF network is used here as a practical mask estimator to investigate the behavior of the different signal representations and masking strategies, rather than as an object of architectural comparison.

Two mask application strategies are considered using the FT-JNF mask estimator. In the first strategy, a single FT-JNF network is trained using channel 0 as the reference channel. The resulting mask is then applied to all channels of the multichannel signal, yielding a shared mask across channels. This approach uses a single network and therefore imposes the same time-frequency attenuation pattern on all channels.

In the second strategy, a separate FT-JNF network is trained for each reference channel.
Each network receives the full multichannel input but is optimized to reconstruct its corresponding reference channel. The resulting channel-specific masks are applied independently to their respective channels, yielding a diagonal masking matrix as in Eqs.~\eqref{eq:mask_mic}, \eqref{eq:mask_amb}, and \eqref{eq:mask_Beam}. This strategy allows the estimated mask to adapt to the characteristics of each channel. Both masking strategies are evaluated for all three signal representations.

In the original FT-JNF architecture, the first BLSTM layer
uses 256 hidden units and the second 128 hidden units,
resulting in approximately 1.2M trainable parameters. In this
work, we reduce the network size by approximately 10 times, using 80 and 40 hidden units in the first and second BLSTM layers, respectively, resulting in approximately 122K parameters, to simulate low-compute and more realistic edge-device applications.

For the beamformer and microphone domains the enhanced signals are transformed back to the SH domain using Eqs.~\eqref{eq:encoder_Beam} and~\eqref{eq:encoder_mic} before binaural decoding. Binaural signals are finally computed via Eq.~\eqref{eq:Binaural} using the Cologne HRTF compilation of the Neumann KU-100~\cite{bernschutz2013spherical}.

\begin{table}[t]
\centering
\caption{Details of the simulation parameters}
\label{tab:simulation_data}
\setlength{\tabcolsep}{4pt}
\begin{tabular}{>{\raggedright\arraybackslash}p{2.0cm} >{\raggedright\arraybackslash}p{5.8cm}}
\toprule
\textbf{Variable} & \textbf{Description} \\
\midrule
Speech & WSJ0 utterances~\cite{paul1992design}, $f_s=16$~kHz \\
\midrule
Room & Rectangular; width $\in[10,20]$, depth $\in[6,10]$, height $\in[3.2,4]$~m;
absorption $\in[0.15,0.5]$ \\
\midrule
Receiver & Random position, at least 1 m away from
the walls, height $1.5$~m \\
\midrule
Target & Front ($0^\circ$), distance $\in[1,3]$~m \\
\midrule
Interferer & Same distance as target; height$\sim\mathcal{N}(1.6, 0.08)$~m; azimuth offset: train $\sim\mathcal{U}[30^\circ,330^\circ]$, test $\in\{+60^\circ,-100^\circ\}$ \\
\midrule
SNR & Target--interferer $\sim\mathcal{U}[-20,10]$~dB; target--noise $=20$~dB \\
\midrule
Noise & Babble source in a random room corner \\
\bottomrule
\end{tabular}
\end{table}

\subsection{Results}

The results of the simulation study are presented in this section. We first examine channel-specific masking based on  first-order Ambisonics representations, followed by shared masking, higher-order representations, and per-channel SI-SDR analysis.

\begin{table*}[t]
\centering
\caption{Performance of the channel-specific IRM and of the FT-JNF estimated masks applied in the microphone (TFM), Ambisonics (TFA), and beamformer (TFB) domains, for first-order Ambisonics ($N=1$). The estimated masks are evaluated using a separate mask for each channel and a shared mask across channels.}
\label{tab:network_mask_comparison}
\setlength{\tabcolsep}{4pt}
\begin{tabular}{llcccccccc}
\toprule
\textbf{Mask} & \textbf{Method}
& \shortstack{\textbf{SI-SDR} \\ (dB) $\uparrow$}
& \shortstack{\textbf{DBSTOI} \\ (-) $\uparrow$}
& \shortstack{\textbf{PESQ} \\ (-) $\uparrow$}
& \shortstack{\textbf{$\Delta$ITD}$_{\mathrm{tar}}$ \\ ($\mu$s) $\downarrow$}
& \shortstack{\textbf{$\Delta$ILD}$_{\mathrm{tar}}$ \\ (dB) $\downarrow$}
& \shortstack{\textbf{$\Delta$ITD}$_{\mathrm{int}}$ \\ ($\mu$s) $\downarrow$}
& \shortstack{\textbf{$\Delta$ILD}$_{\mathrm{int}}$ \\ (dB) $\downarrow$}
& \shortstack{\textbf{$\Delta$C50} \\ (dB) $\downarrow$} \\
\midrule
-- & Noisy & -9.37 & 0.478 & 1.191 & -- & -- & -- & -- & -- \\
\midrule

\multirow{3}{*}{\shortstack[l]{IRM \\ (channel-specific)}}
& TFM   & 3.02 & 0.797 & 2.367 & 0.0 & 0.87 & \textbf{59.4} & 5.32 & 1.75 \\
& TFA   & 8.04 & 0.838 & 2.701 & 0.6 & \textbf{0.07} & 86.9 & \textbf{0.84} & \textbf{1.29} \\
& TFB   & \textbf{10.34} & \textbf{0.852} & \textbf{2.969} & 0.0 & 0.80 & 117.5 & 1.46 & 1.65 \\
\midrule

\multirow{3}{*}{\shortstack[l]{FT-JNF \\ (channel-specific)}}
& TFM   & 1.84 & 0.748 & 1.805 & 0.0 & 0.68 & 83.8 & 3.68 & 2.05 \\
& TFA   & 6.08 & 0.786 & 1.966 & 0.0 & \textbf{0.07} & \textbf{69.4} & \textbf{1.16} & \textbf{1.57} \\
& TFB   & \textbf{7.21} & \textbf{0.791} & \textbf{2.039} & 0.0 & 0.75 & 103.8 & 1.34 & 1.81 \\
\midrule

\multirow{3}{*}{\shortstack[l]{FT-JNF \\ (shared)}}
& TFM   & 1.47 & 0.719 & 1.717 & 0.0 & 0.35 & 82.5 & 3.56 & 2.05 \\
& TFA   & \textbf{5.20} & \textbf{0.773} & \textbf{1.859} & 0.0 & 0.07 & 53.8 & 0.40 & 1.55 \\
& TFB   & 3.17 & 0.757 & 1.668 & 0.0 & \textbf{0.05} & \textbf{43.1} & \textbf{0.36} & \textbf{1.51} \\

\bottomrule
\end{tabular}
\end{table*}

\begin{table*}[t]
\centering
\caption{Performance of the IRM applied in the microphone (TFM), Ambisonics (TFA), and beamformer (TFB) domains, for first-order ($N=1$) and third-order ($N=3$) spherical harmonics representations. The $N=1$ rows are reproduced from Table~\ref{tab:network_mask_comparison} for convenience.}
\label{tab:irm_order_comparison}
\setlength{\tabcolsep}{4pt}
\begin{tabular}{llcccccccc}
\toprule
\textbf{Order} & \textbf{Method}
& \shortstack{\textbf{SI-SDR} \\ (dB) $\uparrow$}
& \shortstack{\textbf{DBSTOI} \\ (-) $\uparrow$}
& \shortstack{\textbf{PESQ} \\ (-) $\uparrow$}
& \shortstack{\textbf{$\Delta$ITD}$_{\mathrm{tar}}$ \\ ($\mu$s) $\downarrow$}
& \shortstack{\textbf{$\Delta$ILD}$_{\mathrm{tar}}$ \\ (dB) $\downarrow$}
& \shortstack{\textbf{$\Delta$ITD}$_{\mathrm{int}}$ \\ ($\mu$s) $\downarrow$}
& \shortstack{\textbf{$\Delta$ILD}$_{\mathrm{int}}$ \\ (dB) $\downarrow$}
& \shortstack{\textbf{$\Delta$C50} \\ (dB) $\downarrow$} \\
\midrule
\multirow{4}{*}{$N=1$}
& Noisy & -9.37 & 0.478 & 1.191 & -- & -- & -- & -- & -- \\
& TFM   & 3.02 & 0.797 & 2.367 & 0.0 & 0.87 & 59.4 & 5.32 & 1.75 \\
& TFA   & 8.04 & 0.838 & 2.701 & 0.6 & 0.07 & 86.9 & \textbf{0.84} & 1.29 \\
& TFB   & \textbf{10.34} & \textbf{0.852} & \textbf{2.969} & 0.0 & 0.80 & 117.5 & 1.46 & 1.65 \\
\midrule
\multirow{4}{*}{$N=3$}
& Noisy & -7.60 & 0.513 & 1.182 & -- & -- & -- & -- & -- \\
& TFM   & 2.06 & 0.803 & 2.295 & 0.0 & 0.89 & 321.9 & 15.47 & 1.30 \\
& TFA   & 9.27 & 0.844 & 2.770 & 0.0 & 0.08 & \textbf{85.0} & \textbf{2.72} & 1.31 \\
& TFB   & \textbf{13.18} & \textbf{0.884} & \textbf{3.360} & 0.0 & 0.35 & 294.4 & 6.89 & 1.45 \\
\bottomrule
\end{tabular}
\end{table*}

\begin{table*}[t]
\centering
\caption{Per-channel SI-SDR (dB) for the noisy signal and after applying the FT-JNF mask, evaluated in the domain in which the mask is applied. ``All'' is the average over the four channels.}
\label{tab:per_channel_sisdr}
\setlength{\tabcolsep}{4pt}
\begin{tabular}{llccccc}
\toprule
& & \multicolumn{5}{c}{\textbf{SI-SDR (dB)}} \\
\cmidrule(lr){3-7}
\textbf{Domain} & \textbf{Masking} & \textbf{ch0} & \textbf{ch1} & \textbf{ch2} & \textbf{ch3} & \textbf{All} \\
\midrule
\multirow{3}{*}{Ambisonics}
& Noisy       & $-5.86$ & $-6.03$ & $-5.98$  & $-6.03$ & $-5.97$ \\
& Shared      & $8.13$  & $6.49$  & $6.65$   & $6.49$  & $6.94$  \\
& Channel-Specific & $8.13$  & $7.99$  & $7.43$   & $8.05$  & $7.90$  \\
\midrule
\multirow{3}{*}{Beamformer}
& Noisy       & $1.80$  & $-9.48$ & $-12.83$ & $-8.43$ & $-7.24$ \\
& Shared      & $10.87$ & $2.57$  & $-0.68$  & $3.58$  & $4.08$  \\
& Channel-Specific & $10.87$ & $4.90$  & $4.45$   & $5.50$  & $6.43$  \\
\midrule
\multirow{3}{*}{Microphone}
& Noisy       & $-4.36$ & $-6.37$ & $-7.19$  & $-6.21$ & $-6.03$ \\
& Shared      & $7.61$  & $6.21$  & $5.57$   & $6.32$  & $6.43$  \\
& Channel-Specific & $7.61$  & $6.76$  & $5.93$   & $6.66$  & $6.74$  \\
\bottomrule
\end{tabular}
\end{table*}

\subsubsection{First-Order Channel-Specific Masking: Oracle and Estimated Masks}
\label{sec:baseline_results}

We first consider the basic configuration, based on first-order Ambisonics representations with channel-specific masking. Two masks are compared: the oracle IRM, which is computed independently for each channel from the known desired and undesired components and the FT-JNF mask, estimated separately for each channel from the noisy multichannel input. This comparison establishes the relationship between oracle and estimated masking before the effects of masking strategy and spatial resolution are examined in the following sections.

As shown in Table~\ref{tab:network_mask_comparison} (upper two parts), the beamformer domain provides the best speech quality and intelligibility performance for both masks, achieving the highest SI-SDR, DBSTOI, and PESQ scores. The Ambisonics domain provides slightly lower performance, while the microphone domain provides the lowest speech enhancement performance. Thus, for this basic configuration, channel-specific masking in the beamformer domain provides the strongest speech enhancement, irrespective of whether the mask is estimated or oracle. This can be explained by the directional selectivity of the beamforming pre-processing, providing an initial enhancement of the target signal. For the estimated mask, speech enhancement performance was lower than for the IRM in all three domains, with lower SI-SDR, PESQ and DBSTOI. While previous work shows the capability of estimated masks to surpass that of IRM \cite{luo2019conv}, here the network is a scaled-down version of FT-JNF with approximately ten times fewer parameters than the original, which may explain the superiority of the IRM. It should be nevertheless noted that the comparison presented here is intended to examine the relative behavior of the different representations under practical mask estimation rather than aiming for best achievable performance.

For the spatial cues, both masks show similar behavior. For the target speaker, $\Delta\mathrm{ITD}$ and $\Delta\mathrm{ILD}$ remain consistently below the JND thresholds of approximately $100~\mu$s for ITD and $1$~dB for ILD~\cite{spencer2016relating,blauert2013technology}. For the interferer, the $\Delta\mathrm{ITD}$ values for both the microphone and Ambisonics representations remain below the JND threshold for both masks, while the beamformer representation is slightly above the threshold. Among the three representations, Ambisonics exhibits the lowest $\Delta\mathrm{ILD}$ error for the interferer, remaining below the JND for the IRM and only slightly above it for the estimated mask, whereas the microphone representation exhibits the highest error.

The preservation of reverberation, measured by $\Delta C_{50}$, is relatively similar across the three representations and follows the same ordering for both masks. Ambisonics provides the best preservation, followed by the beamformer and microphone domains. Mask estimation increases $\Delta C_{50}$ by approximately $0.15$--$0.30$~dB in all three domains compared to IRM. For all representations and both masks, $\Delta C_{50}$ is slightly above the JND, which is approximately 1 dB~\cite{christensen2013iso}, indicating a perceptible but overall limited modification of the reverberant characteristics of the target signal.

Overall, the basic first-order configuration reveals a trade-off between speech enhancement and spatial preservation. Channel-specific masking in the beamformer domain provides the strongest speech enhancement, whereas Ambisonics provides better preservation of the spatial characteristics of the residual interference and reverberation. Importantly, this trade-off, and the relative ordering of the three representations, is preserved when moving from the oracle IRM to the estimated FT-JNF mask, with estimation reducing the absolute performance in all domains without altering the relations between them. This indicates that the main relative differences between the signal representations observed with oracle masking are largely preserved under practical mask estimation. This motivates the use of the IRM in Sec.~\ref{sec:order_results}, where third-order Ambisonics is considered. Using the oracle IRM avoids the need to train separate mask-estimation networks for the larger number of channels introduced by the higher-order representation, while providing a consistent basis for comparing the effect of spatial resolution.

\subsubsection{Effect of Shared Masking}
\label{sec:network_mask_results}

We next investigate how using a single mask across all channels affects the enhancement and spatial preservation compared with channel-specific masking. The results in Table~\ref{tab:network_mask_comparison} (lower part, denoted ``Shared'') show that the shared mask has only a small effect on speech enhancement in the microphone and Ambisonics domains. In contrast, the beamformer domain shows a more pronounced reduction in speech quality and intelligibility. This indicates that the beamformer channels are more sensitive to the use of a shared mask, as their high directional selectivity results in larger differences between channels.

Shared masking consistently reduces the spatial-cue errors of the residual interference across all three representations. The reduction is particularly pronounced for Ambisonics and beamforming, where both $\Delta$ITD and $\Delta$ILD decrease substantially relative to channel-specific masking, indicating improved preservation of the spatial characteristics of the residual interference.  A likely explanation is that applying the same time-frequency attenuation to all channels introduces fewer channel-dependent spectral modifications, thereby reducing distortions of the relative inter-channel relationships. Consequently, the relative spatial relationships between channels are better preserved.

A similar but smaller improvement is observed for reverberation preservation. The $\Delta C_{50}$ error is reduced by shared masking for the Ambisonics and beamformer representations, while remaining essentially unchanged for the microphone representation.

Overall, shared masking provides better preservation of spatial information at the cost of reduced speech enhancement, with this trade-off being most pronounced in the beamformer domain. In contrast, the Ambisonics representation maintains speech enhancement performance close to that of channel-specific masking while benefiting from improved spatial-cue preservation.

\subsubsection{Effect of Ambisonics Order}
\label{sec:order_results}

We next examine the effect of increasing the spatial representation from first to third-order using the oracle IRM, with the results presented in Table~\ref{tab:irm_order_comparison}. Increasing the order substantially increases the difference between the Ambisonics and beamformer domains in terms of speech quality and intelligibility. This is attributed to the increased spatial resolution, which produces more directive beamformers, and denser spatial sampling which enables stronger spatial separation of the target and interfering sources before masking. In contrast, the first-order representation provides broader beamformer responses and therefore limits this spatial selectivity.

The effect of Ambisonics order is also reflected in the preservation of the spatial characteristics of the residual interference. With third-order representations, the difference between the Ambisonics and beamformer domains is more pronounced, with Ambisonics providing better preservation of the interference binaural cues. When the representation is reduced to first-order, the spatial-cue errors of the beamformer become closer to those of Ambisonics. The microphone representation also shows improved preservation of the interference cues at first-order, although its ILD error remains larger than those of the Ambisonics and beamformer representations.

Overall, increasing the Ambisonics order improves the spatial resolution of the beamformer representation and consequently increases its advantage in speech enhancement and its disadvantage in preserving the spatial cues of the interferer. At first-order, the reduced beamformer directivity narrows the performance gap between beamforming and Ambisonics.

\subsubsection{Per-Channel Performance Analysis}
\label{sec:per_channel_sisdr}

Table~\ref{tab:per_channel_sisdr} provides a per-channel analysis of the SI-SDR obtained with the shared and channel-specific FT-JNF mask architectures. Note that first-order Ambisonics is used here so only the FT-JNF masks are presented and not IRM as the former represents more practical values. The results reveal substantial differences between the three signal representations, highlighting the importance of this type of analysis.

For the beamformer representation, the choice of mask type has a strong effect on performance. When the shared mask trained for channel 0 is applied to the other channels, the SI-SDR decreases substantially. In contrast, with the channel-specific mask for which a separate network is trained for each channel, the discrepancy is not as large. This behavior can be explained by the directional nature of the beamformer channels. Since each beamformer is steered toward a different direction, the target-to-interference ratio can vary substantially between channels. 

A different behavior is observed for the microphone representation. The SI-SDR values obtained with the shared and channel-specific masks are generally similar across the four channels, although small differences can be observed. The relatively small differences indicate that the microphone channels have similar, but not identical, signal characteristics, also reflected in the similar SI-SDR values of the noisy channels. This can be attributed to the relatively small distances between microphones in this rather compact spherical microphone array, with the radius of $0.042$~m.

For the Ambisonics representation, a different behavior is observed. Although the four channels have markedly different directivity patterns, the SI-SDR obtained with the shared mask is only slightly lower than that obtained with the channel-specific masks. This is also reflected in the SI-SDR values of the noisy channels, which are similar across the four Ambisonic channels. A possible explanation is that the zeroth-order Ambisonics channel represents the omnidirectional component of the sound field, and therefore its estimated mask is less tied to a particular spatial direction, while all other channels although having a directional response, it is only phase-dependent and not gain-dependent along the azimuth due to the complex exponential term of the spherical harmonics \cite{rafaely2015fundamentals}.

\section{CONCLUSIONS}
\label{sec:Conclusions}
This work investigated time-frequency masking for speech enhancement in spatial audio, comparing microphone, Ambisonics, and beamformer signal representations. Both oracle ideal ratio masks and masks estimated using a lightweight FT-JNF network were considered, and performance was evaluated in terms of speech enhancement, intelligibility, perceptual quality, reverberation preservation, and binaural spatial-cue preservation.

The results show that reducing the Ambisonics representation from third to first-order substantially reduces the performance gap between the Ambisonics and beamformer domains. The reduced spatial resolution limits the directivity of the beamformers, thereby reducing their ability to spatially separate the target and interference. Nevertheless, with channel-specific neural masks, the beamformer domain still provides the highest speech quality and intelligibility, while the Ambisonics domain provides better preservation of the spatial characteristics of the residual interference and reverberation levels.

The comparison of mask-generation strategies further demonstrates an important trade-off between enhancement performance and spatial fidelity. Channel-specific masks generally improve speech enhancement, particularly in the beamformer domain. In contrast, a shared mask can better preserve inter-channel relationships and, in some cases, the reverberant characteristics of the target signal. Notably, in the first-order Ambisonics domain, the shared-mask strategy achieves speech enhancement performance close to that of channel-specific masking while using only a single network. This makes shared masking in the Ambisonics domain an attractive compromise between computational complexity, speech enhancement, and spatial preservation.

The per-channel analysis further shows that the suitability of a shared mask depends strongly on the signal representation. Beamformer channels exhibit substantially different directional characteristics and therefore benefit from channel-specific masks. Microphone channels are more similar, resulting in only modest differences between the two masking strategies. Ambisonics channels can have substantially different spatial characteristics while still allowing a mask optimized for the omnidirectional zeroth-order channel to generalize effectively across channels. These results emphasize that the choice of signal representation and masking strategy should be considered jointly when designing spatial speech enhancement systems.

Overall, the findings demonstrate that there is no single representation that simultaneously optimizes all performance criteria. Instead, the appropriate masking strategy depends on the desired balance between speech enhancement, spatial fidelity, and computational complexity. In particular, first-order Ambisonics with a shared mask provides a practical and efficient solution that maintains strong speech enhancement performance while preserving spatial information, whereas channel-specific beamformer masking is preferable when maximizing speech quality and intelligibility is the primary objective.

\begingroup
\small
\balance
\bibliographystyle{IEEEbib}
\bibliography{refs}
\endgroup

\end{document}